\documentclass[5p,10pt]{elsarticle}
\usepackage{graphics}
\usepackage{amsmath,amsthm,amsfonts,amssymb}
\usepackage{amssymb}
\usepackage{lineno,hyperref}

\begin{document}
\begin{frontmatter}

\title{Sputtering power effects on the electrochromic properties of NiO films}

\author[label3]{Juan R. Abenuz Acu\~na \corref{cor1}}
\cortext[cor1]{Corresponding author}
\ead{sanmaikel23@gmail.com}
\author[label1]{Israel Perez}
\author[label2]{V\'ictor Sosa}
\author[label2]{Fidel Gamboa}
\author[label3]{Jos\'e T. Elizalde}
\author[label3]{Rurik Farias}
\author[label3]{Diana Carrillo}
\author[label3]{Jos\'e L. Enr\'iquez}
\author[label3]{Andr\'es Burrola}
\author[label3]{Manuel Ramos}
\author[label3]{Jesus Netro}
\author[label3]{Pierre Mani}

\address[label1]{National Council of Science and Technology (CONACYT)-Department of Physics and Mathematics, Institute of Engineering and Technology, Universidad Aut\'onoma de Ciudad Ju\'arez, Av. del Charro 450 Col. Romero Partido, C.P. 32310, Ju\'arez, Chihuahua, México}
\address[label2]{Applied Physics Department, CINVESTAV Unidad M\'erida, km 6 Ant. Carretera a Progreso, A.P. 73, C.P. 97310 M\'erida, Yucat\'an, México}
\address[label3]{Department of Physics and Mathematics, Institute of Engineering and Technology, Universidad Aut\'onoma de Ciudad Ju\'arez, Av. del Charro 450 Col. Romero Partido, C.P. 32310, Ju\'arez, Chihuahua, México}

\begin{abstract}
The effect of sputtering power ($P$=60 W-180 W) on the electrochromic properties of nickel oxide films deposited on ITO-coated glass substrates by the RF magnetron sputtering technique was investigated. Crystalline structure and morphology were assessed by X-ray diffraction (XRD) and scanning electron microscopy (SEM), respectively.  The effect of sputtering power on electrochromism of the samples was evaluated with cyclic voltammetry and chronoamperometry. A solution of LiClO$_4$ in propilene carbonate was used for Li insertion/extraction. The chemical properties of the samples before and after Li intercalation were analyzed by X-ray photoelectron spectroscopy (XPS). We observed the cubic phase of NiO with sputtering power mainly affecting crystallinity and grain size. These in turn have an effect on the electrochromic properties. Coloration efficiency reduces from 8.03 cm$^{2}$/C to 3.52 cm$^{2}$/C as sputtering power increases from 60 W to 180 W. XPS analysis reveals that variations of $P$ induce the formation of nickel hydroxides on the film surface. As consequence of changes in crystallinity and morphology the presence of nickel hydroxides increases, showing that not only the electrochromic properties of the samples are affected by the sputtering power but also their chemical properties.
\end{abstract}


\end{frontmatter}

\section{Introduction}
Electrochromic materials are materials capable of changing their optical properties when ions are intercalated or deintercalated as an electric field is applied. Due to this peculiarity, these materials can be exploited to develop electrochromic devices (ECD), controlling in this way, properties such as transmittance and absorbance. Accordingly, some of the applications where these devices can be found are displays, touch screens, smart phones, sunglasses, smart windows, adjustable reflectivity car rear view mirrors, gas sensors and active optical filters among others \cite{gata17a,Moulki2014,R.Wen2015,Q.Liu2017,K.Zhou2017,P.Yang2016}.
An ECD is made of a pile of at least five layers of different materials in a sandwich arrangement. The external layers are transparent conductors (TC) where electric contacts are attached. Next, on each side, two electrodes are placed: one cathodic electrode (CEL) and one anodic electrode (AEL); and in the middle an electrolyte (ELE) is placed as ionic conductor. The applied voltage generates an electric field between the electrodes that makes ions flow from the electrolyte to an electrode. Ions eventually are displaced and captured in the crystalline lattice of one of the electrodes, modifying its optical properties \cite{R.Wen2015,Q.Liu2017, P.Yang2016,Q.Liu2016}.
There are several inorganic compounds that are commonly used in the fabrication of electrodes for ECD. These are transition metal oxides (TMO) where TM is a transition metal such as W, Co, Ni, Ta, Mo, Ir, Ti, V, Mn, Nb, etc. The most studied oxide in ECD is WO$_{3}$ that has one of the largest coloration efficency among the oxides (54.8 cm$^{2}$/A$\cdot$s at 633 nm). On the other hand, more recently, people have turned their attention to NiO \cite{Moulki2014,R.Wen2015,P.Yang2016,Q.Liu2016,X.Song2015,Y.He2019}.
 
Nickel oxide (NiO) is a semiconducting compound crystallizing in either cubic or hexagonal structure. The spatial group for the cubic phase is $Fm\bar{3}m$ with lattice parameter of 4.17 \AA. This material exhibits properties of a $p$-type semiconductor with a band gap ranging from 3.4 eV to 4 eV \cite{hlche05a,rkum15a,ihot04a,sypar10a}. Nickel oxide exhibits electrochromic properties that makes it a good candidate as an anodic material in ECD due to its low cost, good cyclic reversibility and its respectable coloration efficiency ($CE$=42 cm$^{2}$/A$\cdot$s at 550 nm), which is close to that of WO$_{3}$ \cite{Q.Liu2017,P.Yang2016, Q.Liu2016}. 

The growth of NiO films has been investigated using various techniques such as chemical vapor deposition, spin coating, sol-gel method, pulsed laser deposition \cite{Moulki2012,Y.Abe2012}, spray pyrolysis, and radio frequency/direct current sputtering. Radio frequency (RF) sputtering is a convenient technique because offers a better control over the deposition parameters such as thickness, substrate temperature, and transfer of the exact chemical composition \cite{gata17a,aaalg16a}. Previous work has focused on investigating how electrochromic properties are affected by deposition parameters as annealing temperature, oxygen concentration, substrate temperature, etc. \cite{Moulki2014,R.Wen2015,Q.Liu2016,X.Song2015,Y.He2019}. However, to our knowledge there are no works assessing the effect of sputtering power on the electrochromic properties.

In this research we investigate the sputtering  power effects on NiO films grown by RF magnetron sputtering. Crystalline properties and morphology were characterized using X-ray diffraction (XRD) and scanning electron microscopy (SEM) respectively. In order to study the chemical properties before and after Li intercalation X-ray photoelectron spectroscopy (XPS) measurements were performed. Electrochromic properties are evaluated with cyclic voltammetry and chronoamperometry. The present work is valuable to establish the sputtering power as a parameter for manipulating the electrochromic properties of NiO-based ECD.

\section{Experimental}

\subsection{Film growth and annealing}
Six nickel films with a thickness of 200 nm were deposited by RF magnetron sputtering. From these, three were grown on ITO/glass substrates at different sputtering power, that is: $P$=(60, 140, 180) W and so were labeled S60, S140, and S180, respectively. ITO/glass substrates were acquired from MTI corporation with an area of 3 cm $\times$ 1 cm. The thickness of the glass layer is 2.2 mm and that one of ITO is about 200 nm. The three remaining films were grown on Si(100) substrates at the same three sputtering powers. To obtain the NiO phase, the six films were exposed to a postdeposition heat treatment inside the vacumm chamber for 1 hour at 200$^{\circ}$C under an oxygen atmosphere (99.95\% purity) at 1010 mbar. Table 1 gives a summary of the deposition parameters. The films deposited on Si substrates were just used to check sample crystallinity and the others were used for the rest of the study.
\begin{table}[b!]
\centering
\begin{tabular}{c|c} \hline
 $P$/W & 60, 140, 180 \\ \hline
  Target & Ni (99.99\%) \\ \hline
 Substrate type & Si(100) and ITO/glass  \\ \hline
$T_s$/$^\circ$C  & 25 \\ \hline
 $P_B$/mbar & $1.3\times10^{-4}$ \\ \hline
 $P_{\text{Ar}}$/mbar & $1.3\times10^{-2}$\\ \hline
 $P_{\text{O}}$/mbar  & 1010  \\ \hline
\end{tabular}
\caption{Deposition parameters for the films: sputtering power ($P$), target type, substrate type, substrate temperature ($T_s$), base pressure ($P_B$), Ar partial pressure ($P_{\text{Ar}}$) and O partial pressure ($P_{\text{O}}$).}
\label{tab1}
\end{table}

\subsection{Characterization}
To study the crystalline properties, films were characterized using a diffractometer Siemens model D-5000 with Cu K alpha radiation ($\lambda_{0}$=1.540 \AA). XRD patterns were taken at steps of 0.02$^\circ$ with a time per step of 3 s in a Bragg-Brentano configuration, with a voltage of 34 kV and a current of 25 $\mu$A. For morphology study we used a scanning electron microscope Hitachi su5000. The images of the films deposited at $P$=60 W were taken with a magnification of 500 kX and voltage of 15 kV. The films deposited at higher $P$ were measured with a magnification of 100 kX and voltage of 20 kV. To verify the effect of sputtering power on electrochromism of our samples, cyclic voltametry (CV) and chronoamperometry (CA) measurements were carried out using  a CorrTest CS350 electrochemical station in a typical three electrode arrangement. Films were used as working electrodes, and the set was completed with a platinum counter electrode and an Ag/AgCl reference electrode. A cubic optoelectrochemical cell (125 cm$^{3}$) was filled with an electrolyte of LiClO$_{4}$ in propylene carbonate (1 M). Transmittance was measured at $\lambda_{0}$=637 nm, using a high sensitivity light sensor PASCO (model CI-6604) connected to acquisition data system PASCO (model UI-5000). CV measurements were performed between -3 V and 3 V with initial voltage $V_{0}$=3 V and scan rate of 100 mV/s during 30 cycles. CA curves were obtained between -3 V and 3 V ($V_{0}$=3 V) and width step equal to 10 s during 10 steps. To asses the chemical properties of the samples, XPS measurements of the Ni $2p$  and O $1s$ core-levels were carried out using a Thermo Fisher Scientific K-alpha XPS spectrometer. Spectra were generated by monochromatic K-alpha radiation (1486.6 eV) with 30$^{\circ}$ of incident angle between the sample and the X-ray beam. Chemical properties were measured before and after Li intercalation. Resulting spectra were analyzed using CasaXPS software (version 2.3.19PR1.0). A line shape (Gaussian 70\% - Lorentzian 30\%) defined as GL(30) was used for each component and a standard Tougaard background.

\section{Results and discussion}
\subsection{Crystalline Structure}
\begin{figure}[t!]
\begin{center}
\includegraphics[trim=0cm 5cm 2cm 0cm,clip,width=10cm]{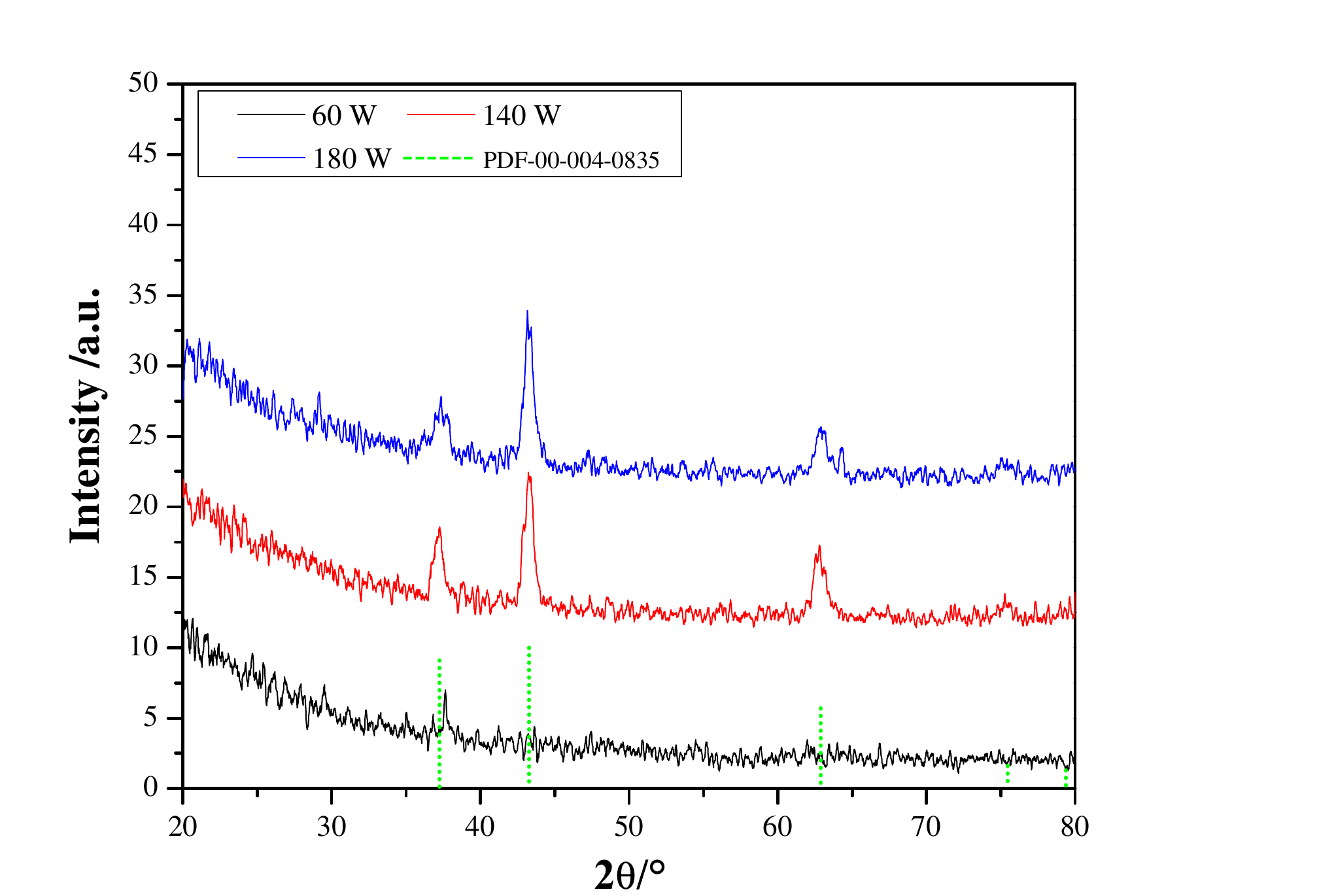}
\caption{Diffraction pattern for the films deposited at different sputtering powers. Green bars at the bottom correspond to the reference PDF 00-004-0835.}
\label{fig1}
\end{center}
\end{figure}
Samples grown on Si substrates exhibit a diffraction pattern with well localized peaks demonstrating the growth of a crystalline phase (see figure \ref{fig1}). The patterns can be indexed to the cubic nickel oxide phase (PDF 00-004-0835), with a spatial group $Fm\bar{3}m$, main diffraction peaks at $2\theta=$ 37$^\circ$, 43$^\circ$, 62$^\circ$, and 75$^\circ$; and lattice parameter  $a=4.17$ \AA. In the case of S60, diffraction peaks corresponding to the crystallographic planes (111) appear with low intensity, an indicative of a low crystallinity. We believe that this is due to the low energy of most of the Ni atoms ejected from the target. As the low energy atoms arrived at the substrate surface they do not have enough energy to be trapped and only those with enough energy are captured and reside in lattice sites, favoring an amorphous phase of Ni \cite{amovchan69a,jathornton74a,nkaiser02a,aanders10a}. 

During the annealing process vacancies are generated and oxygen is adsorbed from the environment tending to form the cubic phase of NiO, however given the amorphous nature of Ni layers, a longer annealing time is required for the atoms to diffuse and order periodically thus manifesting low crystallinity. This is reasonable, for increasing $P$, most atoms from the target have enough energy to be captured by the substrate and reside in preferential sites, this in turn promotes, during annealing, the crystal ordering around these sites and thus inducing the appearance of diffraction peaks  corresponding to crystallographic planes (111), (200), (220), and (311) as can be seen in the rest of the films. All these films showed a preferential orientation along the (200) plane. The crystallite size ($D_c$) of the samples was estimated for the most intense peak according to Scherrer equation:
\begin{equation}
\label{sche}
D_c=\frac{K\lambda_0}{\beta \cos \theta}
\end{equation}
where $K=0.9$, $\lambda_0=1.540\; \text{\AA}$, and $\beta$ is the full width at half maximum of a peak. Results obtained for crystallite size were 119 nm, 11 nm and 15 nm for S60, S140 and S180 respectively.

\subsection{Morphology}
\begin{figure}[t!]
\begin{center}
\includegraphics[width=7cm]{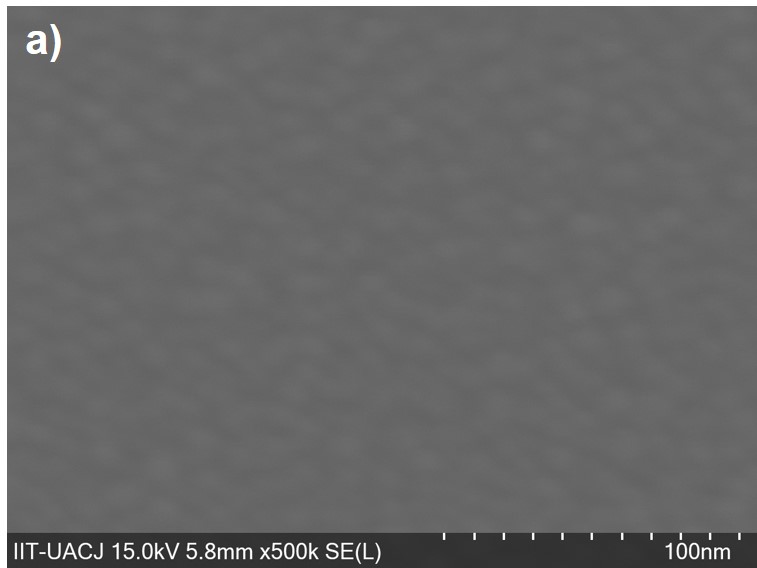} \includegraphics[width=7cm]{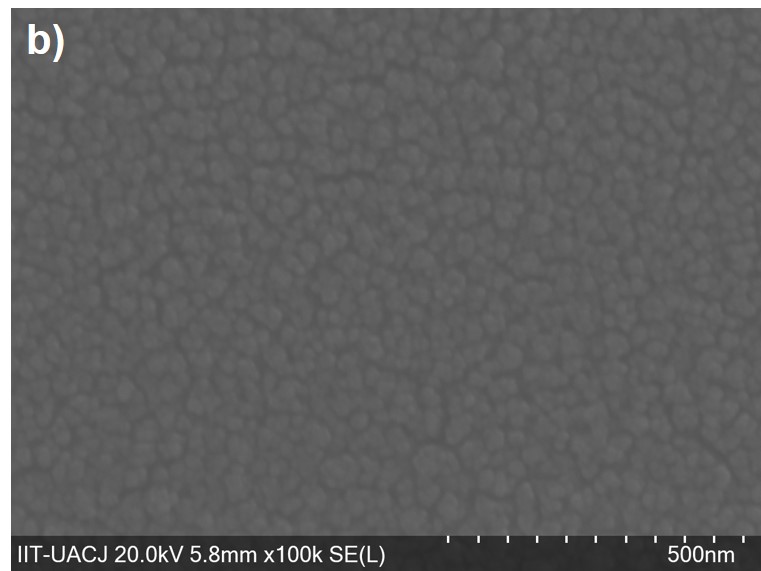}
\includegraphics[width=7cm]{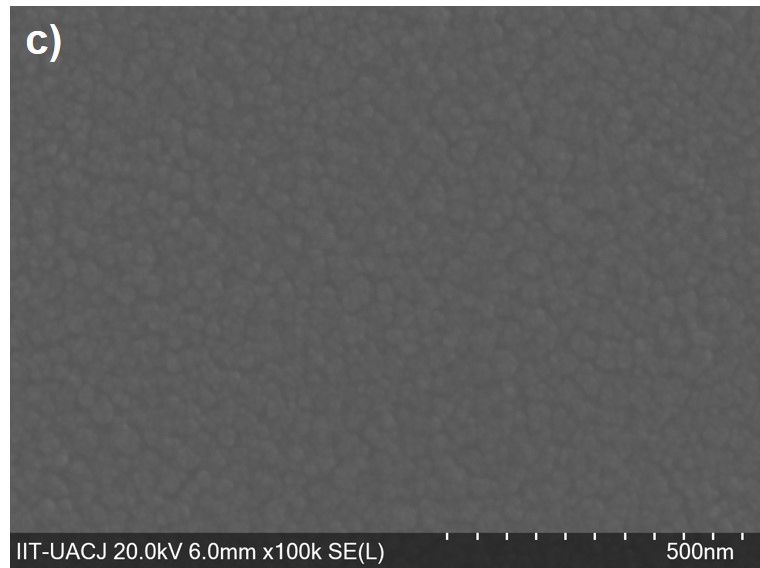} 
\caption{Morphology for films a) S60, b) S140 c)  S180. See text for details.}
\label{fig2}
\end{center}
\end{figure}

\begin{figure}[t!]
\begin{center}
\includegraphics[trim=0cm 4.9cm 2cm 0cm,clip,width=10cm]{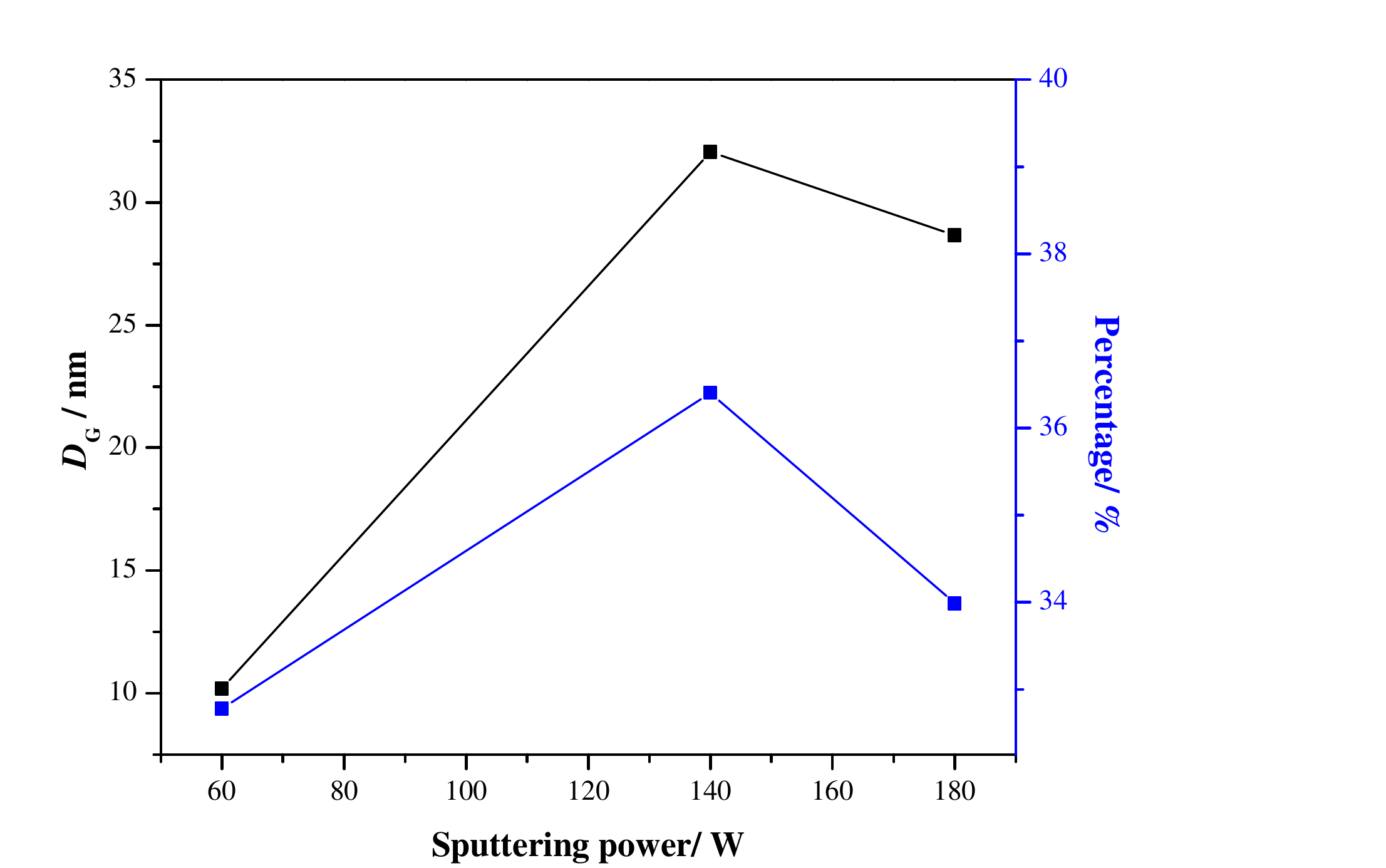}
\caption{Average grain size (black) and percentage of area occupied by grain boundaries (blue) as a function of sputtering power.}
\label{fig3}
\end{center}
\end{figure}

The morphology of the films are shown in figure \ref{fig2}. The analysis reveals the presence of a smooth granular composition, with diluted roughness for S60. A smooth surface is expected for low sputtering powers since deposition takes place with low energy atoms. In the case of S140 and S180, the morphology is also granular but coarse, with marked grain boundaries. A. Ahmed et al. grew several NiO films at 200 W with RF sputtering technique and reported a morphology similar to ours \cite{aaahm17a}. The images of figure \ref{fig2} were used to estimate the average grain size and the percentage of area occupied by the grain boundaries as a function of $P$ (see figure \ref{fig3}). The values of the grain size ($D_{G}$) for the different sputtering powers were (10.1$\pm$0.1) nm, (32.1$\pm$0.8) nm and (28.7$\pm$0.6) nm respectively.
We also calculated the area occupied by grain boundaries for the different sputtering powers. The values were (32$\pm$0.6)\%, (36$\pm$0.3)\% and (33$\pm$0.3)\% respectively (see blue curve in figure \ref{fig3}). By comparison we see that the curves show a similar trend and therefore one can correlate both quantities. The lowest value in the area occupied by grain boundaries of sample S60 can be therefore related to the regular shape that the grains exhibit and the absence of cracks on the surface. On the other hand, increment in the area occupied by grain boundaries for the others samples can be associated to the morphology showed for the samples S140 and S180 with remarkable cracks on surface \cite{aaahm17a}.

\subsection{Electrochromic properties}
\begin{figure*}[t!]
\begin{center}
\includegraphics[trim=0cm 6cm 2cm 0cm,clip,width=17cm]{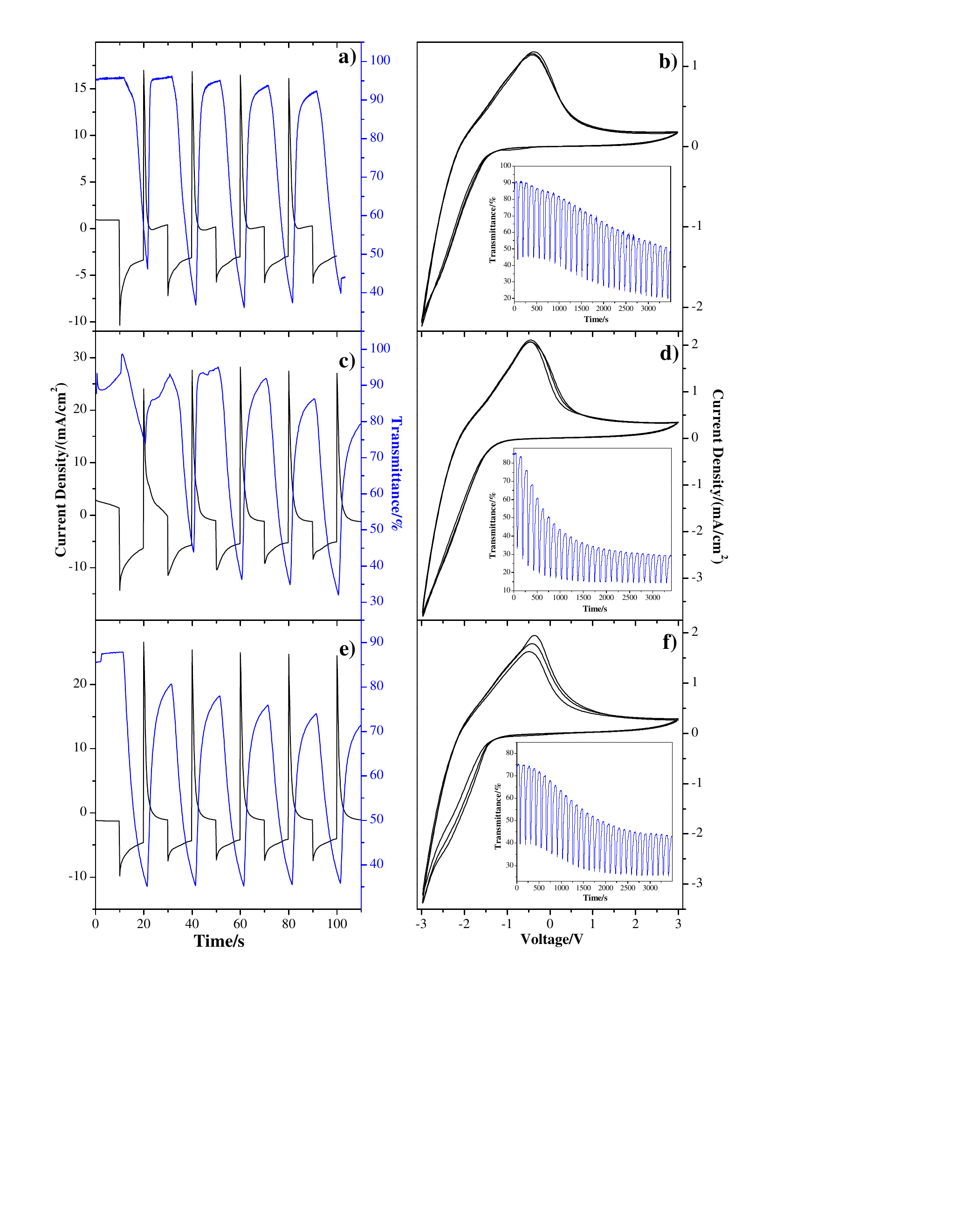}
\caption{Chronoamperometry and transmittance for S60 (a), S140 (c), and S180 (e); and cyclic voltammetry and transmittance (insets) for S60 (b), S140 (d), and S180 (f).}
\label{fig5}
\end{center}
\end{figure*}

In order to study the effect of sputtering power on the electrochromic properties of NiO films chronoamperometry (CA), cyclic voltammetry (CV) and transmittance measurements were performed for all samples. Curves are shown in figure \ref{fig5}. CA and transmittance for S60, S140 and S180 (figure \ref{fig5}a, c and e respectively) show that coloring process shows up during negative current density and the bleached process for positive current density \cite{Q.Liu2017}. Coloration efficiency ($CE$) is the most important property for electrochromic materials and it is defined in equation \ref{CE}:
\begin{equation}\label{CE}
CE=\left| \frac{\Delta OD}{\Delta Q} \right|.
\end{equation}
$CE$ depends on change in optical density $\Delta OD$ determined as
\begin{equation}\label{Optical-density}
\Delta OD= \ln \left(\frac{T_{b}(\lambda_{0})}{T_{c}(\lambda_{0})} \right),
\end{equation}
and the inserted or extracted charge $\Delta Q$. $T_{b}$ and $T_{c}$ are the transmittance in bleached and colored  states at wavelength $\lambda_{0}$.

\begin{table}[b!]
\centering
\resizebox{7 cm}{!} {
\begin{tabular}{ccccc} \hline
Sample &$\frac{\Delta T}{\%}$&\textbf{$\Delta OD$}&$\frac{\Delta Q}{\textbf{C} \cdot \textbf{cm}^{-2}}$& $\frac{CE}{\textbf{cm}^{2} \cdot \textbf{C}^{-1}}$ \\
&&&&\\ \hline
S60     &  60.3     &  0.99          & -0.12   &  8.02   \\ 
S140    &   51.4    &  0.92          & -0.19   &  4.81   \\ 
S180    &   35.7   &  0.64          & -0.18   &  3.52   \\ 
\end{tabular}
}
\caption{Electrocromic properties for the samples: Optical modulation ($\Delta T$), change in optical density ($\Delta OD$), charge difference ($\Delta Q$) and coloration efficiency ($CE$).}
\label{tab2}
\end{table}
Coloration efficiencies are 8.02 cm$^{2}$/C, 4.81 cm$^{2}$/C and 3.52 cm$^{2}$/C for S60, S140, and S180, respectively (table \ref{tab2}). This indicates that sputtering power affects the electrochromic properties, reducing the $CE$ as $P$ increases. As we have found above, such changes can be attributed to variations in cristallinity and morphology (figures \ref{fig1} and \ref{fig3}). According to Kailing Zhou et al. amorphous NiO films with a further improvement in cristallinity exhibits higher values of $CE$ as result of large amounts of active sites for electrolyte diffusion. Consequently an enhancement in crystal ordering generates the decrement of $CE$ \cite{K.Zhou2017} as occurs with our results.  Possibly the improvement in crystallinity with $P$ intensifies the variations in lattice stress and rises the formation of defects as result of Li insertion/extraction that reduces the absorption and transmission of light. Complementary, the increase in average grain size and area occupied by grain boundaries as $P$ increases (figures \ref{fig2} and \ref{fig3}) contribute to high stress and a large expansion/contraction of grains due to Li intercalation that provokes the mechanical deterioration of the film \cite{K.Zhou2017,Q.Liu2016} affecting the change in transmittance and $CE$.

CV measurements for all samples are shown in figures \ref{fig5}b, d and f respectively, only cycles 10, 20 and 30 were plotted. Inset figures display the transmittance for all cycles. In the loops for S60 and S140 (figure \ref{fig5}b and d) we observe that the current density at oxidation peak remains stable around 1.18 mA/cm$^{2}$ at -0.38 V and 2.06 mA/cm$^{2}$ at -0.46 V respectively. The graph for S180 (figure \ref{fig5}f) exhibits an increment from 1.62 mA/cm$^{2}$ to 1.94 mA/cm$^{2}$ as the number of cycles rises, suggesting a high reaction activity \cite{Xia2008}. Reduction peak was analyzed and S60 shows the peak around 0.043 mA/cm$^{2}$ at -0.98 V just for cycle 10. The increase in current density beyond the reduction peak for all samples is probably related to a high electron conductivity in the colored state and internal electronic leakage arising from defects such as metallic particles, pinholes and dust, as a consequence the current density is composed of kinetic transfer of Li$^{+}$ ions, reversely kinetic transfer of the charge-balancing electrons and leakage-inducing electrons \cite{Q.Liu2017}. Cycles 20 and 30 reveal the absence of the reduction peak suggesting a higher contribution from leakage current. S140 does not exhibit a reduction peak in any cycle while S180 and S60 exhibit a similar behavior in the 10 cycle with a current density peak at -0.77 V. The transmittance and therefore the optical modulation decreases as the number of cycles evolve.

The maximum transmittance dependence of the sputtering power after the first cycle was analyzed for the three samples. We observe a decrement of the transmittance from 90.8\% at 60 W to 75\% at 180 W; indicating that higher sputtering powers reduced the maximum transmittance. Since we have determined that higher sputtering power favors the crystallinity of films, larger grain boundaries and larger grains, this seems to affect also the Li intercalation, the absorption of light and thus the transmittance. Electrochromic measurements suggest that electrochromic properties of NiO changes by varying the sputtering power $P$.  

\subsection{Chemical analysis}
\begin{figure*}[t!]
\begin{center}
\includegraphics[width=6cm]{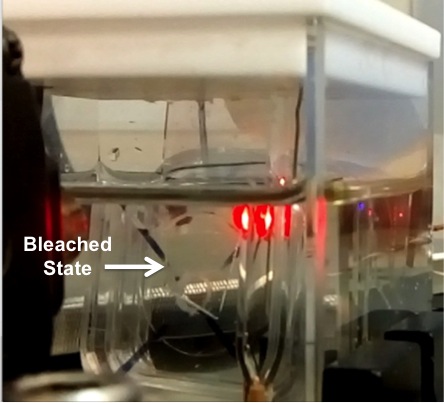}\hspace{2cm} \includegraphics[width=6cm]{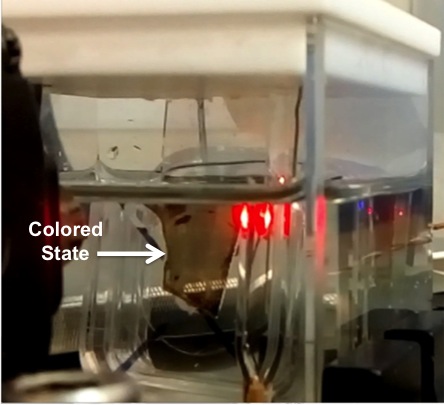}
\caption{(Color online) S60 dipped in the electrolite solution in the bleached (left) and colored (right) states at a bias of $-3$ V}
\label{bleached1}
\end{center}
\end{figure*}

XPS measurements were performed for all samples before and after Li intercalation to study the effect of sputtering power on the chemical properties. Li intercalation was attained by dipping the sample in the electrolite and applying a bias of $-3$ V for 1 s, thus leaving the samples in the colored state (see figure \ref{bleached1}). 

\begin{figure}[b!]
\begin{center}
\includegraphics[trim=0.5cm 6cm 2cm 0cm,clip,width=10cm]{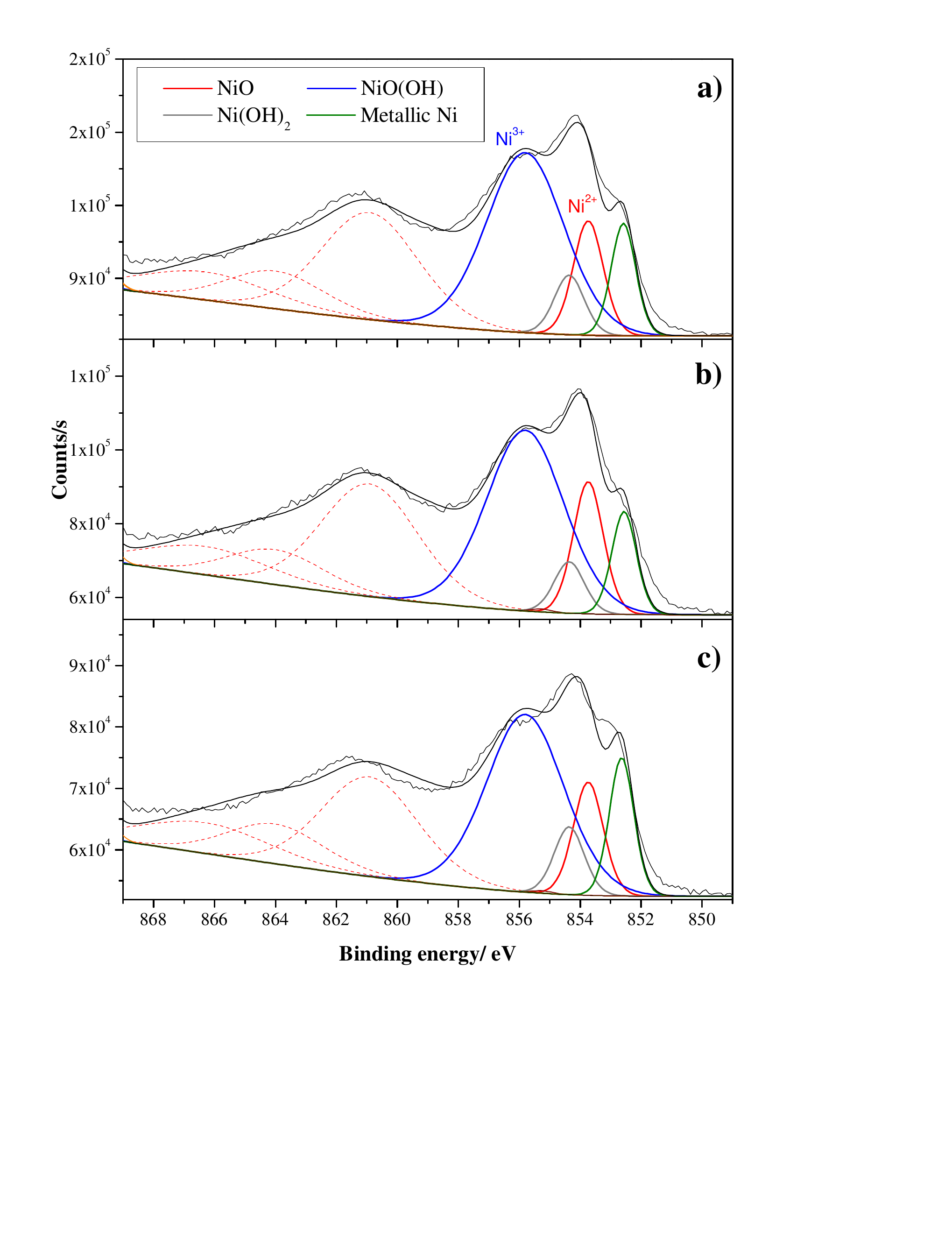}
\caption{XPS spectra for Ni $2p$ core-level of a) S60, b) S140 and c) S180 before Li intercalation.}
\label{fig4a}
\end{center}
\end{figure}

\begin{figure}[b!]
\begin{center}
\includegraphics[trim=0.5cm 6cm 2cm 0cm,clip,width=10cm]{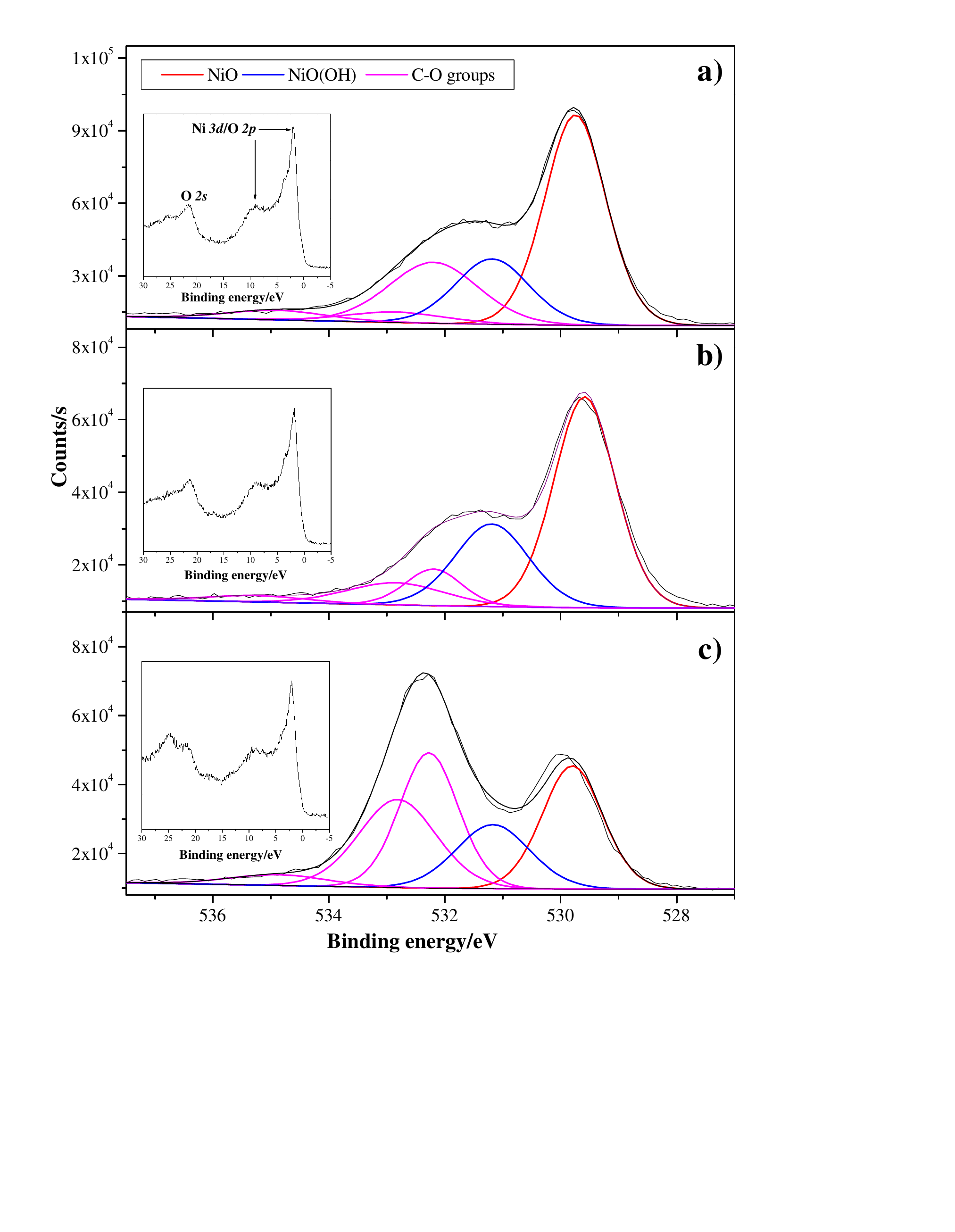}
\caption{XPS spectra for O $1s$ core-level of a) S60, b) S140 and c) S180 before the Li intercalation. Inset show the valence band.}
\label{fig4b}
\end{center}
\end{figure}

The spectra for Ni $2p$ and O $1s$ core-levels of S60, S140 and S180 before Li intercalation are presented in the figures \ref{fig4a} and \ref{fig4b}, respectively. A peak at 853.7 eV in the Ni spectra is observed, corresponding to Ni$^{2+}$ for NiO phase. Peaks at $~$854.3 eV  and  $~$855.8 eV are attributed to Ni(OH)$_{2}$ and NiO(OH) (oxidation state Ni$^{2+}$ and Ni$^{3+}$, respectively) \cite{mtya13a,I. Saric2016,B. P. Payne2012,B. P. Payne2009,M. C. Biesinger2009}. 
According to Rui-Tao et al. \cite{R.Wen2015}, OH$^{-}$ groups can be adsorbed on the surface of NiO films when there are small amounts of water. The presence of both NiO(OH) and Ni(OH)$_{2}$ in our samples can be associated to water adsorbed from the environment during film manipulation. 
We also analyzed the C $1s$ core-level (not shown) and observed carbon contamination which is adsorbed mainly during film manipulation, suggesting that the water comes also from the environment. The peaks at 854.3 eV and  855.8 eV suggest an admixture of NiO(OH)/Ni(OH)$_{2}$ on the surface, however, the Ni(OH)$_{2}$ signal is lower than the signal of NiO(OH), indicating a higher presence of NiO(OH) \cite{B. P. Payne2009,Y. Abe2012,A. Van Der Ven2006}. One peak at $~$852.5 eV is observed for all samples that is due to the presence of metallic Ni \cite{B. P. Payne2009,M. C. Biesinger2009}. Perhaps, its presence is associated to reduction of nickel atoms present in different nickel compounds as a result of the annealing process, especially at the surface \cite{S. Oswald2004}. Peaks observed beyond 860 eV (red dash lines) are satellites. 

\begin{figure}[t!]
\begin{center}
\includegraphics[trim=0.5cm 6cm 2cm 0cm,clip,width=10cm]{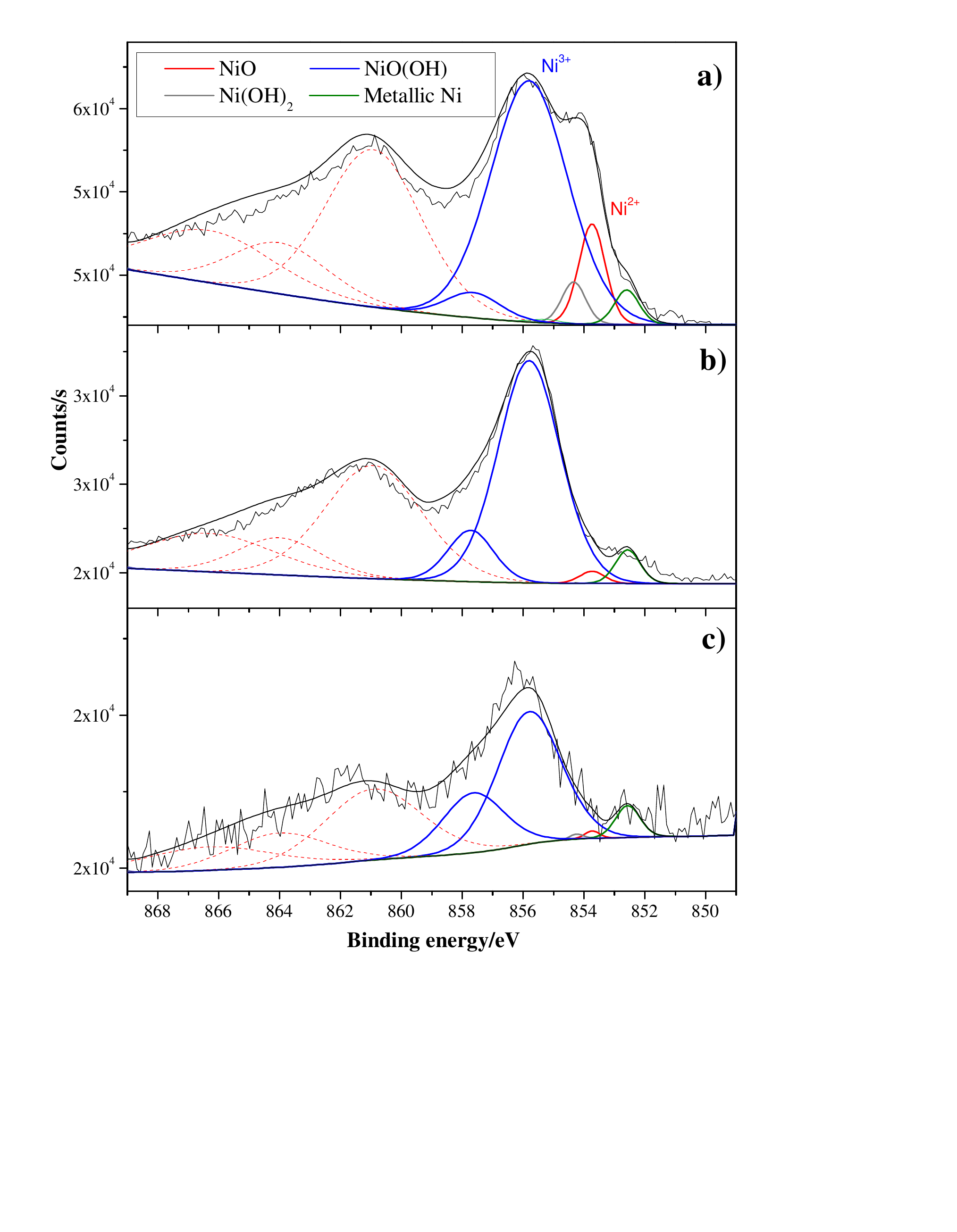}
\caption{XPS spectra for Ni $2p$ core-level of a) S60, b) S140 and c) S180 after the Li intercalation.}
\label{fig6a}
\end{center}
\end{figure}
\begin{figure}[t!]
\begin{center}
\includegraphics[trim=0.5cm 6cm 2cm 0cm,clip,width=10cm]{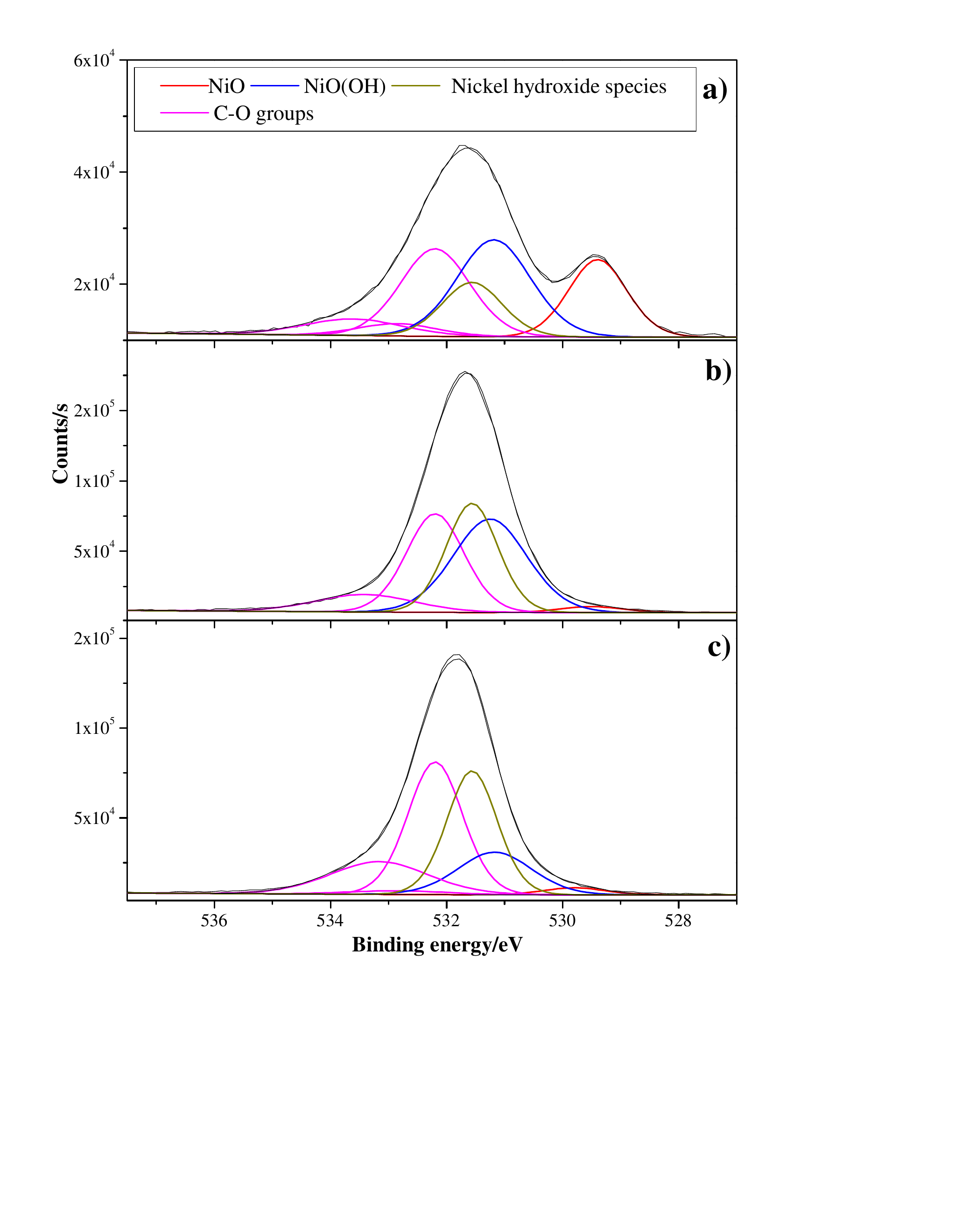}
\caption{XPS spectra for O $1s$ core-level of a) S60, b) S140 and c) S180 after the Li intercalation.}
\label{fig6b}
\end{center}
\end{figure}

The O $1s$ core-level are presented in figure \ref{fig4b}. A low binding energy peak at $~$529.7 eV corresponding to NiO phase appears. The peak at ~531.2 eV can be attributed to NiO(OH) \cite{mtya13a,I. Saric2016,B. P. Payne2012,B. P. Payne2009,M. C. Biesinger2009}. Peaks observed at higher binding energies ($~$532.2 eV) can be related to organic contamination such as carbon which is ubiquitous in most samples and is adsorbed easily when the samples are exposed to the environment \cite{B. P. Payne2009}. 
To confirm the presence of NiO phase, we also studied the valence band
(insets in figure \ref{fig4b}). The typical spectra corresponding to NiO is observed for all samples, with peaks at (1.9, 8.9) eV associated to hybridization of Ni $3d$ and O $2p$ core-levels. Another peak appears at 21.3 eV corresponding to O $2s$ core-level. These results suggests a high presence of NiO and a hydroxilation of the surface. To determine the oxygen and nickel concentration for NiO phase in our samples, we computed the oxygen to nickel ratio (O/Ni) by estimating the area under the peaks of Ni $2p_{3/2}$ and O $1s$ core-levels, respectively. The ratios were 1.33, 1.15, and 1.39 for S60, S140 and S180, respectively. Accordingly, these results imply the formation of nickel hydroxides at the surface and the nonstoichiometry of the samples, suggesting that sputtering power affects the oxygen and nickel concentration.Previous studies have reported an O/Ni ratio between 1.5 and 2 for both NiO(OH) and Ni(OH)$_{2}$ \cite{B. P. Payne2012}.
  
To study the effect of $P$ on the chemical composition and their relationship with changes in electrochromic properties XPS spectra were  analyzed after Li intercalation. The spectra for Ni $2p$ and O $1s$ core-levels for all samples in the colored state are shown in figures \ref{fig6a} and \ref{fig6b} respectively. For the Ni $2p$ core-level we identify the same peaks as in the previous case. In addition to these peaks, another signal is spotted at 857.6 eV that can be associated to nickel hydroxides species \cite{B. P. Payne2012,B. P. Payne2009,M. C. Biesinger2009}, possibly NiO(OH) or Ni(OH)$_{2}$. We observe that the Ni$^{2+}$ signal diminishes as $P$ increases while the signal attributed to nickel hydroxide species increases. The spectra of the O $1s$ core-level (figure \ref{fig6b}) show the same behavior as in the previous case, however a new peak at 531.5 eV, related to nickel hydroxides \cite{B. P. Payne2012,B. P. Payne2009,M. C. Biesinger2009,A.Carley1983}, appears for all samples. It is evident that the signal associated to NiO (Ni$^{2+}$) decreases as $P$ increases while the signal related to nickel hydroxides at 531.5 eV increases. We thus assume that hydroxylation of surface is taking place at higher sputtering powers. Our findings strongly indicate that, after interacting with OH$^{-}$ ions, NiO reversibly turns into NiO(OH) and from NiO(OH)$_{2}$ to NiO(OH) \cite{R.Wen2015,Y.Abe2012,A.Stenman2013}. The existence of this nickel species on the film surface before Li intercalation (figure \ref{fig4a} and \ref{fig4b}) in addition to the presence of OH$^{-}$ groups in the electrolyte as a result of small amounts of water, apparently contributes to chemical mechanisms that facilitate the appearance of nickel hydroxides. Moreover, as result of changes in the morphology and crystallinity with changes in $P$, large quantities of Li ions may be inserted and extracted from the film, rising the number of chemical reactions that promote the hydroxilation of the film. 

According to the preceding discussion, one realizes that before Li intercalation, XPS analysis reveals the presence of both NiO and nickel hydroxides on the surface. Also we observe an effect of the sputtering power on the stoichiometry of NiO. After Li intercalation, we observe the same oxidation states as in the previous case, however the atomic concentration of the Ni hydroxides largely differs from that before Li intercalation. Therefore, it is clear that changes in crystallinity and morphology driven by variations in sputtering power have an impact on the formation of Ni hydroxides. 

\section{Conclusions}
We deposited nickel oxide films at different sputtering powers (60 W-180 W). Both film crystallinity and morphology were affected by the changes in sputtering power. The XRD characterization showed diffraction patterns corresponding to cubic NiO, manifesting a change in the preferential orientation of the films, from the (111) crystallographic plane to the (200) plane as $P$ increases. We found a close relationship between the average grain size and the area occupied by grain boundaries as a function of $P$. Both quantities follow a similar tendency. S60 show the highest value of $CE$ and it drops for S140 and S180 respectively, it can be correlated to decrement in optical density as $P$ increases. For CV measurements the maximum transmittance in bleached states after the first cycle also reduces as $P$ increases. CA and CV measurements indicate that as $P$ increases $CE$ diminishes. Chemical composition determined with XPS technique before Li intercalation show the presence of several phases such as NiO, NiO(OH), Ni(OH)$_{2}$ and metallic Ni. Valence band and the O/Ni ratio for all samples suggest that films are mainly composed of NiO and the partial hydroxilation of surface. After Li intercalation chemical composition exhibit a relationship with $P$ indicating that hydroxylation of the surface is promoted at higher sputtering power as a result of a possible increase in Li insertion in the film lattice stimulated by the different morphologies and crystallinities of the samples. 

Results obtained from this research invariably indicate that crystalline properties, morphology, chemical composition and electrochromic properties of NiO films deposited by RF sputtering can be altered trough sputtering power variations.

\section*{Acknowledgements}
We are grateful to Dra. Claudia Rodr\'iguez, Daniel Aguilar, Hortencia Reyes, for their technical support during the UV-VIS, XRD, and SEM sessions. The authors gratefully acknowledge the support from the National Council of Science and Technology (CONACYT) México and the program C\'atedras CONACYT through project 3035. We also acknowledge the support from the Universidad Aut\'onoma de Ciudad Ju\'arez through program PIVA 334-18-12.

\bibliographystyle{elsarticle-num}

\end{document}